\documentclass[a4paper,11pt]{article}
\usepackage{amstext,amssymb}
\usepackage{amsmath}
\usepackage{graphicx}
\usepackage{dcolumn}
\usepackage[hyperfootnotes=false]{hyperref}
\usepackage{xspace}
\usepackage{braket}
\usepackage{color}
\usepackage{units}
\usepackage{slashed} 
\usepackage{mathrsfs}  
\usepackage{cite}
\usepackage{bm}

\def \epsilon {\varepsilon} 

\newcommand{\gbl}{g_{BL}}
\newcommand{\mzp}{M_{Z^\prime}}
\newcommand{\mhone}{M_{H_1}}
\newcommand{\mhtwo}{M_{H_2}}
\newcommand{\mdm}{M_\mathrm{DM}}

\title{\bf A new $\boldsymbol{B-L}$ model without right-handed neutrinos}
\author{Sudhanwa Patra\footnote{sudha.astro@gmail.com}\\
\textit{\small Center of Excellence in Theoretical and Mathematical Sciences}\\
\textit{\small Siksha 'O' Anusandhan University, Bhubaneswar-751030, India}
\and
Werner Rodejohann\footnote{werner.rodejohann@mpi-hd.mpg.de},~~Carlos E. Yaguna\footnote{carlos.yaguna@mpi-hd.mpg.de}\\
\textit{\small Max-Planck-Institut f\"ur Kernphysik}\\
\textit{\small Saupfercheckweg 1, 69117 Heidelberg, Germany}}
\date{}
\begin{document}

\maketitle

\begin{abstract}
\noindent
We propose and study a novel extension of the Standard Model based on the $B-L$ gauge symmetry that can account for dark matter and neutrino masses. In this model, right-handed  neutrinos are absent and the gauge  anomalies are canceled instead by four chiral fermions with fractional $B-L$ charges. After the breaking of  $U(1)_{B-L}$, these fermions arrange themselves into two  Dirac particles, the lightest of which is automatically stable and plays the role of the dark matter.  We determine the regions of the parameter space consistent with the observed dark matter density and show that they can be partially probed via direct and indirect dark matter detection or collider searches at  the LHC. Neutrino masses, on the other hand, can be explained by a variant of the type-II seesaw mechanism involving  one of the two scalar fields responsible for the dark matter mass. 
\end{abstract}
%

\maketitle
\newpage

\section{Introduction}
\label{sec:introduction}
Models in which the difference between baryon and lepton number, \mbox{$B-L$}, is gauged are economic and well-motivated extensions of the Standard Model~\cite{Jenkins:1987ue,Buchmuller:1991ce,Khalil:2006yi, Basso:2008iv,Iso:2009ss, Kanemura:2014rpa} that may shed light on the origin of neutrino masses \cite{Petcov:2013poa,Gonzalez-Garcia:2015qrr} and the nature of the dark matter \cite{Bergstrom:2000pn,Bertone:2004pz,Drees:2012ji} -- two of the most pressing problems in particle physics today. Among the possible realizations of such models, the  minimal one is that based on the gauge group $SU(2)_L\times U(1)_Y\times U(1)_{B-L}$, which simply extends the Standard Model (SM) with an extra $U(1)$ of $B-L$.

In these models, the cancellation of gauge anomalies is usually achieved with the addition of three  right-handed neutrinos, which simultaneously allow to explain neutrino masses via the type-I seesaw mechanism \cite{Minkowski:1977sc,Yanagida:1979as,GellMann:1980vs,Mohapatra:1979ia}. Several attempts have also been made to incorporate the dark matter within these scenarios \cite{Rodejohann:2015lca, Lindner:2013awa, Okada:2010wd, Okada:2012sg, Basak:2013cga, Duerr:2015wfa, Guo:2015lxa, Dasgupta:2016odo}. It has been known for some time, though, that the anomalies in this model can also be canceled in other ways. In particular, a model with 3 singlet fermions with $B-L$ charges $5$, $-4$ and $-4$ was first proposed in \cite{Montero:2007cd} and has received some attention lately -- see e.g. \cite{Sanchez-Vega:2014rka,Ma:2014qra,Sanchez-Vega:2015qva,Ma:2015mjd}.

In this paper we present a new $B-L$ gauge model, based on the $U(1)_{B-L}$ extension of the SM, in which the right-handed neutrinos are absent and the gauge anomalies are canceled instead by four  chiral fermions that are singlets under the SM gauge group but have fractional charges under $U(1)_{B-L}$.  These charges forbid any tree level interactions between  the  Standard Model particles and the new fermions,  rendering the lightest of them automatically stable and therefore  a viable dark matter candidate. Two important features  of this model are thus that the fields responsible for anomaly cancellation also explain the dark matter and that the stability of the dark matter particle is automatic -- there is no need to impose any extra discrete symmetries to ensure it. 

Besides these four chiral fermions, the model includes two scalar fields, also singlets of the SM, with $B-L$ charges $1$ and $2$, which spontaneously break the $B-L$ symmetry and give Dirac-type masses to the new fermions. Another scalar field, a triplet of $SU(2)$, is further required  to explain neutrino masses via a variant of the type-II seesaw mechanism \cite{Schechter:1980gr,Magg:1980ut,Mohapatra:1980yp,Lazarides:1980nt}. Interestingly,  the necessary induced vacuum expectation value is here generated by one of the scalar particles responsible for the dark matter mass, thus indirectly connecting neutrino masses and dark matter. 

The plan of the paper is as follows.  In the next section the model is introduced and described in detail. We  write down the full Lagrangian, implement symmetry breaking, and find the fermion and scalar mass matrices. The dark matter phenomenology is presented in section \ref{sec:DM}. Specifically, we determine the regions of the parameter space consistent with the observed value of the dark matter density and  discuss the role of current and planned dark matter experiments in probing them. In section \ref{sec:lhc} the LHC bounds are examined while  in  section \ref{sec:numass} we explain how   neutrino masses are generated within this model. Finally, we summarize our results in section \ref{sec:con}.

\section{A new $U(1)_{B-L}$ gauged model}
\label{sec:model}
The $B-L$ gauge extension of Standard Model (SM), where the difference between baryon and lepton number is defined as a local gauge symmetry, is one of the simplest extensions from the point of view of a self-consistent gauge theory. It naturally appears in well-motivated scenarios for physics beyond the SM, such as left-right theories and unification models. Here, we will focus on a model based on the $SU(3)_C\times SU(2)_L \times U(1)_Y \times U(1)_{B-L}$ gauge symmetry. With just the SM fermions, this model is not anomaly-free as both
\begin{equation}
\mathcal{A}_1\left[U(1)^3_{B-L}\right] = 
\mathcal{A}_2\left[\mbox{(gravity)}^2 \times U(1)_{B-L}\right], 
\end{equation}
are non-zero. The usual way of overcoming this problem is to  add  right-handed neutrinos $N_{Ri}, (i=1,2,3)$, each of which has a $B-L$ charge of $-1$. In addition, these right-handed neutrinos may also explain neutrino masses via a type-I seesaw mechanism. 

In this paper, we would like to propose an alternative way of canceling the gauge anomalies that does not invoke right-handed neutrinos. As we will see, this novel scenario provides a direct connection to dark matter and offers also an interesting link to neutrino masses.

\label{sec:u1BL}
\subsection{Particle content}

In a model without right-handed neutrinos, the $B-L$ gauge anomalies can be canceled instead by the following four chiral fermions
\begin{align}
\xi_L(4/3), \, \eta_L(1/3), \, \chi_{1R}(-2/3), \, \chi_{2R}(-2/3)\, , 
\end{align}
which are singlets under the SM gauge group but have fractional charges under $B-L$ (the number in parenthesis). Here the fields $\xi_L$ and $\eta_{L}$ are left-handed, while $\chi_{iR}$ ($i=1,2$) are right-handed. First of all, let us check that the gauge anomalies indeed vanish
\begin{align}
\hspace*{0cm}&\mathcal{A}_1\left[U(1)^3_{B-L}\right] =\mathcal{A}^{\rm SM}_1\left[U(1)^3_{B-L}\right]+ 
\mathcal{A}^{\rm New}_1\left[U(1)^3_{B-L}\right] ,\nonumber \\
&=-3+\left[(4/3)^{3}+(1/3)^{3}-(-2/3)^{3}-(-2/3)^{3} \right]=0\,\nonumber\\
&\mathcal{A}_2\left[\mbox{(gravity)}^2 \times U(1)_{B-L} \right]=\mathcal{A}^{\rm SM}_2+ \mathcal{A}^{\rm New}_2\nonumber \\ \nonumber
&=-3+\left[(4/3)+(1/3)-(-2/3)-(-2/3) \right]=0\,.
\end{align}
In addition to these fermions, the model includes two new scalars, $\phi_1,\phi_2$, also singlets under the SM, with $B-L$ charges $1$, $2$  respectively, which   break the $B-L$ symmetry  and   give masses, via their vevs, to the new fermions. These fermions arrange themselves into two Dirac particles, the lightest of which is automatically stable -- without the need of ad hoc discrete symmetries -- and constitutes a viable dark matter candidate. Thus, the dark matter is explained in this model by the same fields that are required to cancel the gauge anomalies. Moreover, since the correct relic density is obtained, within the thermal scenario, for dark matter masses around the TeV scale, the $B-L$ breaking scale should also lie close to TeV and, therefore, not far from  the LHC reach. Hence, this scenario predicts a low  $B-L$ breaking scale and could be tested not only via dark matter experiments but also at colliders.   

\begin{table}[tb!]
\begin{center}
\begin{tabular}{|c|c|c|c|c|}
	\hline
		& Field	& $SU(2)_L\times U(1)_Y$	& $U(1)_{B-L}$	\\
	\hline
	\hline
	Fermions& $Q_L \equiv(u, d)^T_L$		& $(\textbf{2}, 1/6)$	& $1/3$	\\
		& $u_R$					& $(\textbf{1}, 2/3)$	& $1/3$	\\
		& $d_R$					& $(\textbf{1}, -1/3)$	&$1/3$	\\
		& $\ell_L \equiv(\nu,~e)^T_L$		& $(\textbf{2}, -1/2)$	&  $-1$	\\
		& $e_R$					    & $(\textbf{1}, -1)$ &  $-1$	\\
        \hline
		& $\xi_{L}$				    & $(\textbf{1}, 0)$	&   $4/3$	\\
		& $\eta_{L}$				& $(\textbf{1}, 0)$	&   $1/3$	\\
		& $\chi_{1R}$				& $(\textbf{1}, 0)$	&   $-2/3$	\\
		& $\chi_{2R}$				& $(\textbf{1}, 0)$	&   $-2/3$	\\
	\hline
	Scalars	& $H$					& $(\textbf{2}, 1/2)$	&   $0$	\\
			& $\phi_{1}$			& $(\textbf{1}, 0)$	&   $1$	\\  
			& $\phi_{2}$			& $(\textbf{1}, 0)$	&   $2$	\\ 
			& $\Delta$			& $(\textbf{3}, 1)$	&   $-2$	\\ 
	\hline
	\hline
\end{tabular}
\caption{\small Particle content of the $U(1)_{B-L}$ model.}
\end{center}
\label{tab:New_BL_DM}
\end{table}

Finally, one more scalar, $\Delta$, triplet of $SU(2)$ and with $B-L=-2$, helps neutrinos to acquire non-zero Majorana masses via a variant of the type-II seesaw mechanism involving also $\phi_2$. Indeed, as explained in section \ref{sec:numass}, the vacuum expectation value of $\Delta$ is induced by the SM Higgs $H$ and the scalar $\phi_2$, thus linking neutrino masses and dark matter within this model.  The complete particle content, with the respective quantum numbers, is presented in Table~\ref{tab:New_BL_DM}.

To explain the smallness of neutrino masses, the scalar field $\Delta$ must be heavy ($M_\Delta \gg 1$ TeV), so it effectively decouples from other phenomena at lower energies. To simplify our analysis, in the following we will include $\Delta$ only in our discussion of neutrino masses, in section \ref{sec:numass}.

\subsection{The  Lagrangian}
  
The most general Lagrangian involving  the new fields and consistent with the $SU(2)_L\times U(1)_Y\times U(1)_{B-L}$ gauge symmetry  is given by
\begin{align}
\label{eq:TheModel}
	\mathscr{L}_\text{BL} 
	&=     i \, \overline{\xi_{L}} \left( \slashed{\partial} + \frac{4}{3} i\,g_\text{BL}  \,Z_\mu^\prime \gamma^\mu \right)\,\xi_{L} 
	     + i \, \overline{\eta_{L}} \left( \slashed{\partial} + \frac{1}{3} i\,g_\text{BL} \,Z_\mu^\prime \gamma^\mu \right)\,\eta_{L} 
	     \nonumber \\
	&~~~ 	
	     +  i \, \overline{\chi}_{iR} \left( \slashed{\partial} - \frac{2}{3} i\,g_\text{BL} \,Z_\mu^\prime \gamma^\mu \right)\,\chi_{iR} 
	\nonumber \\
	&~~~ - \left(\alpha_{i}\, \overline{\xi_L} \chi_{i\,R}\,\phi_2 
     + \beta_{i}\, \overline{\eta_L} \chi_{i\,R}\, \phi_1+ h.c.\, \right) 
	\nonumber \\
	&~~~ + |\left( \partial_\mu +2 \,i\,g_\text{BL}\,Z'_\mu \right) \phi_2|^2
		 + |\left( \partial_\mu + \,i\,g_\text{BL}\,Z'_\mu \right) \phi_1|^2 
	\nonumber \\
	&~~~ - \frac{1}{4} F_{Z^\prime}^{\mu \nu}F^{Z^\prime}_{\mu \nu}
	     - V( H,  \phi_1,  \phi_2 )
	  \, ,
\end{align}
where $\alpha_i$, $\beta_i$ are new Yukawa couplings, $\gbl$ is the gauge coupling associated to the $U(1)_{B-L}$ group, $Z_\mu^\prime$ is its corresponding gauge boson, and   $F_{\mu \nu}^{Z^\prime}$ the respective field strength tensor. The scalar  potential, $V ( H,  \phi_1,  \phi_2 )$, will be discussed in the following subsection.

Notice that bare mass terms for the new fermions are forbidden by the $B-L$ symmetry. Their masses are generated instead from the Yukawa terms once $\phi_{1,2}$ acquire vacuum expectation values.

Remarkably, this Lagrangian  automatically includes an \emph{accidental} $Z_2$ symmetry under which the new fermions are odd while the other fields are even. Thus, the lightest of these fermions will be stable and a viable dark matter candidate. 

\subsection{The scalar sector and symmetry breaking}
The most general scalar potential involving $H$, $\phi_1$ and $\phi_2$ and consistent with the gauge symmetry of our model is
 \begin{align}
V(H,\phi_1,\phi_2)= & \mu^2_H  H^\dagger H + \lambda_H (H^\dagger H)^2 
     + \mu^2_1 \phi^\dagger_1 \phi_1 + \lambda_1 (\phi^\dagger_1 \phi_1)^2 
      +\mu^2_2 \phi^\dagger_2 \phi_2  \nonumber \\
      &+ \lambda_2 (\phi^\dagger_2 \phi_2)^2  +\rho_{1} (H^\dagger H) (\phi^\dagger_1 \phi_1)
      +\rho_{2} (H^\dagger H) (\phi^\dagger_2 \phi_2)\nonumber \\
      &+\lambda_{3} (\phi^\dagger_1 \phi_1) (\phi^\dagger_2 \phi_2) 
      +\mu \left( \phi_2 \phi^{\dagger^2}_1 + \phi_2^\dagger \phi^{^2}_1 \right).
      \label{eq:V}
\end{align} 
The conditions for this potential to be bounded from below read
\begin{equation}
 \lambda_H, \lambda_1, \lambda_2 \geq 0,\;\;
 \rho_1 + \sqrt{\lambda_H \lambda_1} \geq 0\, , \;\; \rho_2 + \sqrt{\lambda_H \lambda_2} \geq 0\, , \;\; 
  \lambda_3 + \sqrt{\lambda_1 \lambda_2} \geq 0 \,.
\end{equation}

The spontaneous symmetry breaking of $SU(2)_L \times U(1)_Y \times U(1)_{B-L}$ down to $SU(2)_L \times U(1)_Y$ is achieved by assigning non-zero vacuum expectation values (vevs) to the scalars $\phi_1$ and $\phi_2$ at a scale above the  electroweak phase transition scale. Later, $SU(2)_L \times U(1)_Y$ breaks down to electromagnetism 
via the neutral component of the Higgs doublet, $H^0$. 

The fields $H^0$, $\phi_1$ and $\phi_2$ can be parametrised in terms of real scalars and pseudoscalars as
\begin{align}
&H^0 =\frac{1}{\sqrt{2} }(v+h)+  \frac{i}{\sqrt{2} } G^0\, , \nonumber \\
& \phi_1 = \frac{1}{\sqrt{2} }(v_1+h_1)+  \frac{i}{\sqrt{2} } A_1\,, \nonumber \\
& \phi_2 = \frac{1}{\sqrt{2} }(v_2+h_2)+  \frac{i}{\sqrt{2} } A_2\, .
\end{align}
with $\langle H^0\rangle=v/\sqrt2$, $\langle \phi_1\rangle=v_1/\sqrt2$, $\langle \phi_2\rangle=v_2/\sqrt2$. The minimisation conditions of the scalar potential imply that
\begin{align}
\mu^2_H& = -\left(\lambda^2_H v^2 + \frac{ \rho_1}{2} v^2_1 +   \frac{ \rho_2}{2} v^2_2 \right),  \nonumber \\
\mu^2_1& = -\left(\lambda^2_1 v^2_1 + \frac{ \rho_1}{2} v^2 +   \frac{ \lambda_3}{2} v^2_2+\sqrt{2} v_2 \mu \right) , \nonumber \\
\mu^2_2& = -\left(\lambda^2_2 v^2_2 + \frac{ \rho_2}{2} v^2 +   \frac{ \lambda_3}{2} v^2_1 +\frac{1}{\sqrt{2}} \frac{v^2_1 \mu}{v_2} \right).  \nonumber 
\end{align}

Because  $\phi_1$ and $\phi_2$ are charged under $B-L$, their vevs induce a non-zero mass for the neutral gauge boson, $Z^\prime$, associated with the $B-L$ gauge symmetry. This mass is given by  
\begin{eqnarray}
	M^2_{Z^\prime}= g^2_{BL} \left( v_1^2+ 2 v_2^2 \right).
\end{eqnarray}
It is convenient to define a new dimensionless parameter, $\tan\beta$, as the ratio between the vevs of the scalars fields $\phi_1$ and $\phi_2$: $\tan\beta =\frac{v_1}{v_2}$. Thus, 
\begin{equation}
M^2_{Z^\prime}= g^2_{BL} v_2^2 \left(2+\tan^2\beta\right)
\end{equation}
so that $v_1$, $v_2$ can be written in terms of $\mzp$, $\gbl$ and $\tan\beta$.

Since the $Z^\prime$ couples to the SM fermions, its mass can be constrained with collider data. From LEP II the bound reads \cite{Carena:2004xs,Cacciapaglia:2006pk}
\begin{equation}
\label{eq:LEPII}
\frac{\mzp}{\gbl}\gtrsim 7~\mathrm{TeV}.
\end{equation}
LHC data also set limits on $\mzp$, as will be discussed in section \ref{sec:lhc}.

\subsection{Scalar masses}
The terms proportional to $\rho_1$ and $\rho_2$ in equation (\ref{eq:V}) induce mixing between the SM Higgs boson and the new scalar fields of this model. Since the scalar boson observed at the LHC with a mass of $M_h = 126\,$GeV is very much SM-like \cite{Khachatryan:2014jba, Aad:2015zhl}, this mixing is necessarily small. For simplicity, in the following we will  neglect this mixing, effectively setting $\rho_{1,2}$ to zero.  

The scalar CP-even spectrum thus consist of the SM Higgs plus two other states which mix with each other  according to the mass matrix
\begin{align}
\label{eq:scalarMass}
\mathcal{M}^2_{\rm Higgs} = 
\begin{pmatrix} 
2 \lambda_1 v^2_1     &   v_1( \lambda_3  v_2 + \sqrt{2}  \mu )   \\    
v_1( \lambda_3  v_2 + \sqrt{2} \mu )   &   2 \lambda_2 v^2_2 -\frac{\mu v^2_1}{\sqrt{2}v_2}
\end{pmatrix}
\end{align}
in the $(h_1,h_2)$ basis. The resulting mass eigenstates, denoted by $H_1$ and $H_2$, are related to $h_{1,2}$ via the mixing angle, $\theta$:
\begin{equation}
\begin{pmatrix} h_1\\h_2\end{pmatrix}=\begin{pmatrix}\cos\theta &\sin\theta\\
-\sin\theta &\cos\theta\end{pmatrix}\begin{pmatrix}H_1\\H_2\end{pmatrix}.
\end{equation}
It is convenient to take as free parameters of the scalar sector the physical masses of $H_{1,2}$ ($M_{H_{1,2}}$) and the mixing angle $\theta$.  The couplings $\lambda_i$ can then be expressed in terms of them as
\begin{align}
\lambda_1=& \frac{1}{2v_1^2}\left[\cos^2\theta\,M^2_{H_1}+\sin^2\theta\,M^2_{H_2}\right],\\
\lambda_2=& \frac{1}{2v_2^2}\left[\sin^2\theta\,M^2_{H_1}+\cos^2\theta\,M^2_{H_2}+\frac{\mu v_1^2}{\sqrt 2v_2}\right],\\
\lambda_3  =&\frac{1}{v_1v_2}\left[\sin\theta\cos\theta(M_{H_2}^2-M_{H_1}^2)-\sqrt 2\mu v_1\right].
\end{align}

The mass matrix for the CP-odd scalars in the basis ($A_1$, $A_2$) is given instead by
\begin{align}
\mathcal{M}^2_{\mbox{\small CP-odd}} = 
\begin{pmatrix} 
-2 \sqrt{2} v_2 \mu                &  \sqrt{2} v_1  \mu      \\
\sqrt{2} v_1 \mu                   &   -\frac{v^2_1}{\sqrt{2}} \frac{\mu}{v_2}  
\end{pmatrix},
\end{align}
and, as expected, has an eigenvalue equal to zero -- the would-be Goldstone boson that becomes the longitudinal mode of the $Z^\prime$. The mixing angle, $\alpha$, in this case is entirely determined by the vevs: $\sin\alpha= \sqrt{4v_2^2/(v_1^2+4v_2^2)}$. The physical CP-odd eigenstate will be denoted by $A$ and its mass by $M_A$. The parameter $\mu$ is then given as
\begin{equation}
\mu = -\frac{M_A^2\,\sin^2\alpha}{2\sqrt 2 v_2}\,.
\end{equation}

This model predicts, therefore, the existence of 3  scalar fields beyond the SM Higgs: $H_1$, $H_2$ and $A$. These fields have scalar interactions among themselves, gauge interactions with the $Z^\prime$, and Yukawa interactions with the new fermions. 

\subsection{Fermion masses}
The  scalar fields $\phi_1$, $\phi_2$ are required to give masses to the new fermions after the spontaneous breaking of the $B-L$ symmetry. From the Lagrangian, we can read the fermion mass matrix  as
%
\begin{align}
\label{eq:diracnu}
	\mathcal{L}_{\rm mass} &=\left(\begin{array}{cc}\overline{\xi_L}& \overline{\eta_{L}}\end{array}\right) 	\left( \begin{array}{cc}  \alpha_1\,\langle \phi_2\rangle & \alpha_2\, \langle \phi_2\rangle       \\
\beta_1\,\langle\phi_1\rangle & \beta_2\langle\phi_1\rangle
		\end{array}
	\right) \left(\begin{array}{c}\chi_{1\,R}\\ \chi_{2\,R} \end{array}\right) + h.c.,
\end{align}
which has a Dirac form. Hence, this model contains two Dirac mass eigenstates, denoted by $\psi_{1,2}$, the lightest of which will be the dark matter particle. They are related to the original gauge eigenstates via the mixing matrices $U_L$ and $U_R$ that diagonalize the fermion mass matrix as
\begin{equation}
\begin{pmatrix}\xi_L\\ \eta_L\end{pmatrix}=U_L\begin{pmatrix}\psi_{1\,L}\\ \psi_{2\,L}\end{pmatrix},\qquad
\begin{pmatrix}\chi_{1\,R}\\ \chi_{2\,R}\end{pmatrix}=U_R\begin{pmatrix}\psi_{1\,R}\\ \psi_{2\,R}\end{pmatrix},
\end{equation}
where $\psi_1=\psi_{1\,L}+\psi_{1\,R}$, $\psi_2=\psi_{2\,L}+\psi_{2\,R}$ and $U_{L,R}$ can each be parametrized by a mixing angle $\theta_{L,R}$ as
\begin{equation}
U_{L,R}=\begin{pmatrix}\cos\theta_{L,R} &\sin\theta_{L,R}\\
-\sin\theta_{L,R} &\cos\theta_{L,R}\end{pmatrix}.
\end{equation}
It is convenient to take as free parameters determining the fermion mass matrix the two mass eigenvalues $M_{1,2}$ and the two mixing angles $\theta_{L,R}$.  The couplings $\alpha_{1,2}$ and $\beta_{1,2}$ can then be written in terms of these  parameters as
\begin{align}
\alpha_1  =& \frac{\sqrt 2}{v_2}\left(\cos\theta_L\cos\theta_R \,M_1+\sin\theta_L\sin\theta_R \,M_2\right),\\
\alpha_2  =& \frac{\sqrt 2}{v_2}\left(-\cos\theta_L\sin\theta_R \,M_1+\sin\theta_L\cos\theta_R \,M_2\right),\\
\beta_1  =& \frac{\sqrt 2}{v_1}\left(-\sin\theta_L\cos\theta_R \,M_1+\cos\theta_L\sin\theta_R \,M_2\right),\\
\beta_2  =& \frac{\sqrt 2}{v_1}\left(\sin\theta_L\sin\theta_R \,M_1+\cos\theta_L\cos\theta_R \,M_2\right).
\end{align}

The interactions terms between the new neutral fermions and the $Z^\prime$ is given, in the mass eigenstate basis, by
\begin{align}
\mathscr{L}_{\psi Z^\prime}=&-\frac{g_{BL}}{3}\bigg[
\overline{\psi_{1}} \gamma^\mu 
\bigg\{\left(3 \cos^2\theta_L+1 \right) P_L- 2 P_R \bigg\} \psi_{1} \nonumber \\
&+\overline{\psi_{2}} \gamma^\mu 
\bigg\{\left(3 \sin^2\theta_L+1 \right) P_L- 2 P_R \bigg\} \psi_{2} \nonumber \\
&+\overline{\psi_{1}} \gamma^\mu 
\left(3 \sin^2\theta_L \cos\theta_L \right)P_L \psi_{2} 
+\overline{\psi_{2}} \gamma^\mu 
\left(3 \sin^2\theta_L \cos\theta_L \right)P_L \psi_{1} 
\bigg]
Z^\prime_\mu\,,
\end{align}
which does not depend on $\theta_R$. Without loss of generality we  assume in the following that $\psi_1$ is lighter than $\psi_2$ and, therefore, the dark matter candidate. The vector ($g_{\psi V}$) and axial ($g_{\psi A}$) couplings of the dark matter particle to the $Z^\prime$, which play a crucial role in the dark matter   phenomenology,  are then given by 
\begin{align}
g_{\psi V} = \frac{g_{BL}}{6} \left(1-3 \cos^2\theta_{L} \right) ,~~g_{\psi A} = \frac{g_{BL}}{2} \left(1+\cos^2\theta_{L} \right).
\end{align}
 
We now have all the ingredients required to quantitatively study the implications of this model.  In the next section, we investigate in detail the dark matter phenomenology. 

\section{Dark Matter Phenomenology}
\label{sec:DM}
\subsection{Thermal relic density}
\begin{figure}[tb!]
\begin{tabular}{ccc}
\includegraphics[scale=0.42]{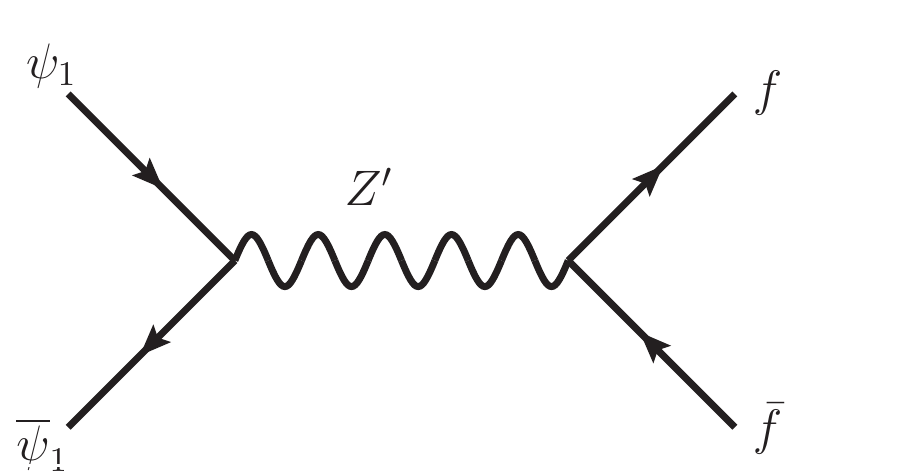} & \includegraphics[scale=0.42]{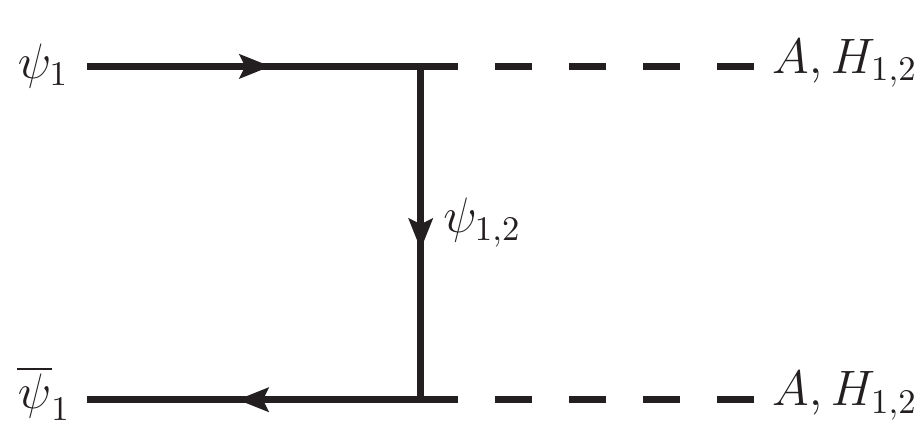} & \includegraphics[scale=0.42]{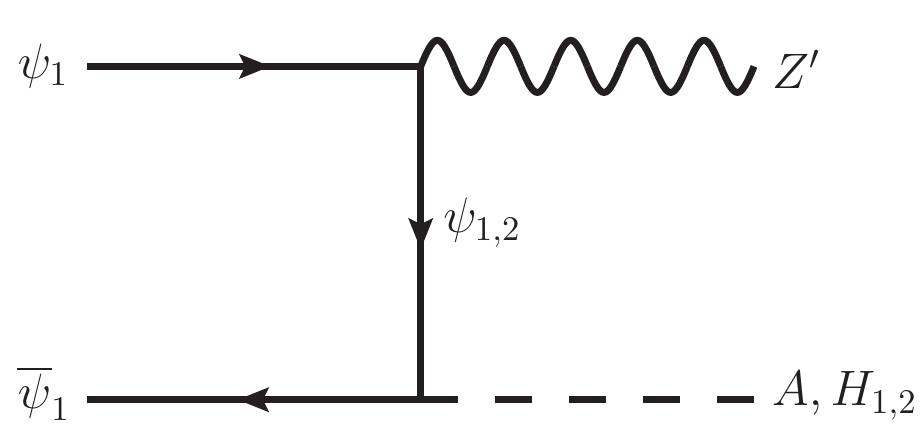} 
\end{tabular}
\caption{\small Some of the diagrams that contribute to dark matter annihilation in this model. \label{fig:diagrams}} 
\end{figure}
Being neutral and stable, the lightest Dirac fermion of this model,  $\psi_1$, is a viable cold dark matter candidate. It has  $U(1)_{B-L}$ gauge interactions mediated by the $Z^\prime$ and also  scalar interactions, induced by the new Yukawa couplings,  with $H_1$, $H_2$ and $A$. Both interactions may contribute to dark matter annihilation in the early Universe. Figure \ref{fig:diagrams} shows some representative Feynman diagrams for dark matter annihilation in this model. The annihilation into SM fermions mediated by the $B-L$ gauge boson (left diagram in figure \ref{fig:diagrams}) has a cross section which, in the non-relativistic limit and neglecting fermion masses, is given by 
\begin{align}
\sigma v \left(\overline{\psi_1} \psi_1 \to Z^{\prime^*} \to \bar{f}f \right)
&=
\frac{N_c^f \mdm^2 g_{\psi V}^2g_{fV}^2}{\pi
\big[ \left(4 M^2_{\rm DM}- M^2_{Z^\prime} \right)^2 + M^2_{Z^\prime} \Gamma^2_{Z^\prime}\big] 
 } \,,
\end{align}
where   $M_{\rm DM}$ denotes the dark matter mass, $\Gamma_{Z^\prime}$ is the total decay width of the $Z^\prime$, and ($N_c^f$, $g_{fV}$) is equal to ($1$, -$\gbl$) for leptons and to ($3$, $\gbl/3$) for quarks. Hence, dark matter annihilation  mediated by  $Z^\prime$ depends only on four free parameters: $\mdm$, $\mzp$, $\gbl$ and $\theta_L$ (via $g_{\psi V}$).

In addition, the dark matter could also annihilate into final states containing scalar particles ($H_{1,2}$, $A$) via several diagrams, two of which are displayed in figure \ref{fig:diagrams}. Besides the scalar masses, these annihilations into scalar particles depend  on other parameters such as $\tan\beta$, $\theta_{L,R}$, and the mixing angle in the scalar sector, $\theta$. For an accurate  calculation of the relic density, we have relied on micrOMEGAs \cite{Belanger:2013oya} (after implementing the model via LanHEP \cite{Semenov:2010qt}), which automatically takes into account all the relevant contributions to the annihilation cross section and properly treats the annihilations close to the resonance. 
\begin{figure}[tb!]
\begin{center}
\includegraphics[scale=0.4]{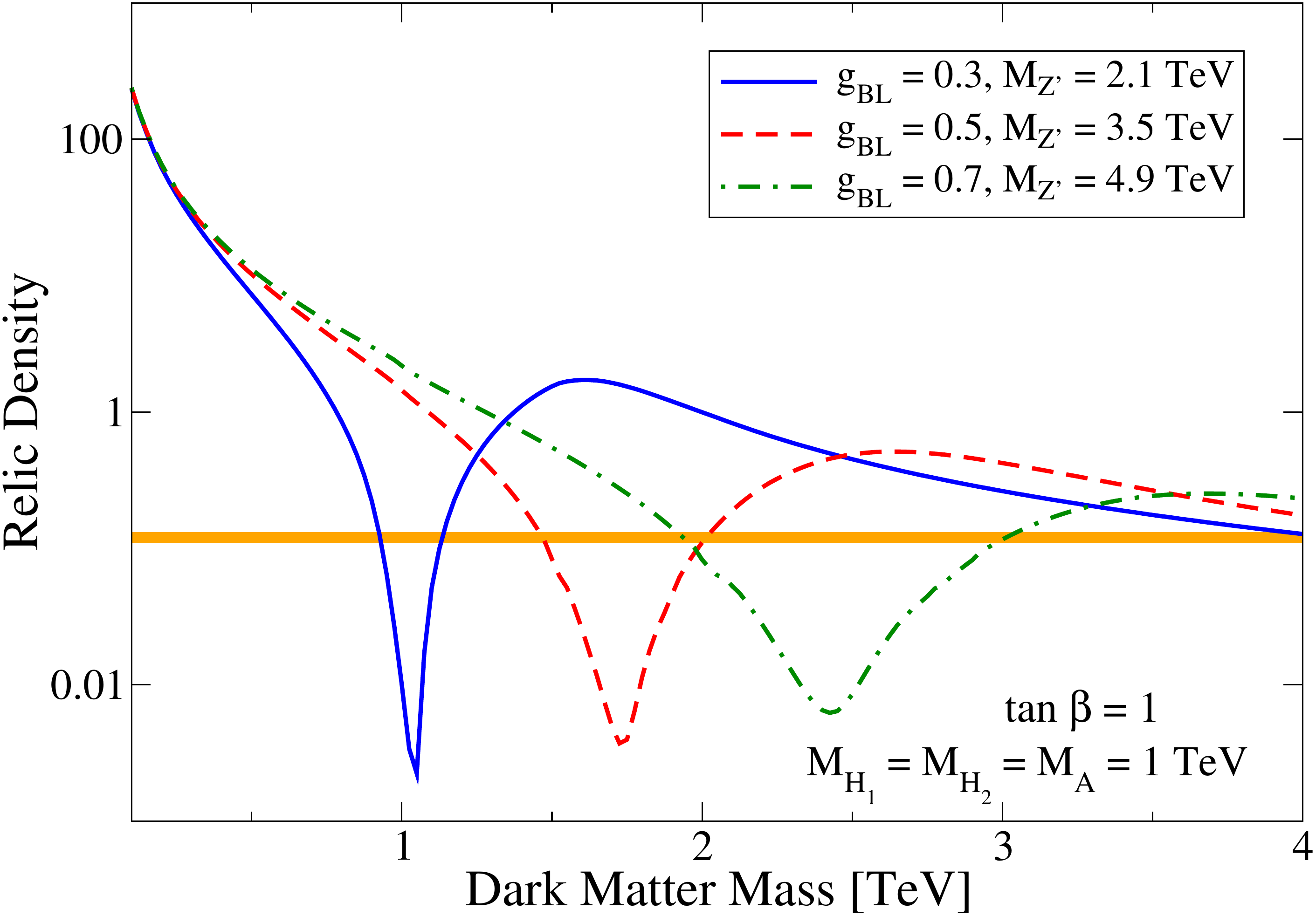}
\caption{\small The dark matter relic density as a function of the DM mass for different values of  ($\gbl$, $\mzp$): ($0.3$, $2.1$ TeV) in  blue, ($0.5$, $7$ TeV) in red, and  ($0.7$, $4.9$ TeV) in green. In this figure, $\tan\beta=1$, $\mhone=\mhtwo=M_A=1$ TeV and all mixing angles -- $\theta_{L,R}$, $\theta$ -- were set to zero. The horizontal orange band corresponds to the region consistent with the observed dark matter density.\label{fig:rd}} 
\end{center}
\end{figure} 

To illustrate the dependence of the dark matter relic density with the parameters of the model, we show in figure \ref{fig:rd} the predicted relic density as a function of the dark matter mass for three different combinations of $(\gbl, \mzp)$, all of them consistent with the LEPII limit from equation (\ref{eq:LEPII}). In this figure, $\tan\beta=1$, the scalar masses were set to 1 TeV,  and the mixing angles in the scalar and fermionic sectors were assumed to be negligible ($\theta=\theta_{L,R}=0$). From the figure we see that the minimum value of the relic density is obtained at the resonance, $\mdm\sim \mzp/2$, and that its value increases with the $Z^\prime$  mass. As expected, this resonance region becomes wider as $\mzp$ increases.  The horizontal orange band corresponds to the region consistent with the observed dark matter density \cite{Ade:2013zuv}. For this set of parameters, the dark matter constraint can be satisfied in two different regions: around the resonance, and for dark matter masses close to $4$ TeV. 

\begin{figure}[tb!]
\begin{center}
\includegraphics[scale=0.4]{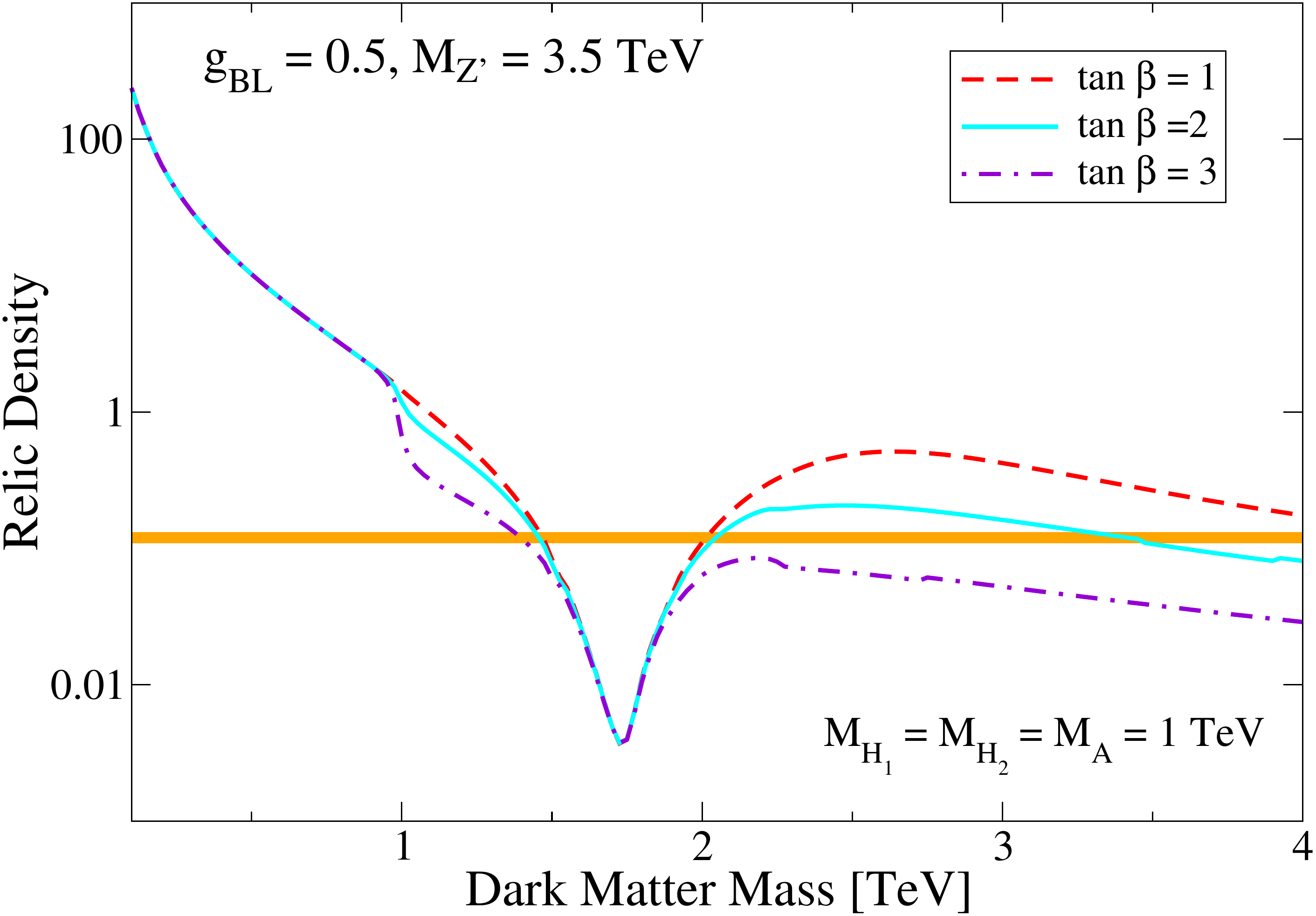}
\caption{\small The dark matter relic density as a function of the DM mass for different values of  $\tan\beta$: $1$ (red), $2$ (cyan) and $3$ (violet). In this figure, $\gbl=0.5$, $\mzp=3.5$ TeV, $\mhone=\mhtwo=M_A=1$ TeV and all mixing angles -- $\theta_{L,R}$, $\theta$ -- were set to zero. The horizontal orange band corresponds to the region consistent with the observed dark matter density.\label{fig:rdtb}} 
\end{center}
\end{figure} 

If we now allow $\tan\beta$ to vary, the picture changes slightly due to the contribution from  the  final states containing scalar particles, as illustrated in figure \ref{fig:rdtb}. In it, the $B-L$ gauge coupling and gauge boson mass were fixed -- respectively at $0.5$ and $3.5$ TeV -- but three different values of $\tan\beta$ were considered: $1$ (red), $2$ (cyan) and $3$ (violet). Since $\tan\beta$ modifies only the couplings to the  scalars, the relic density does not depend on its value when the dark matter mass is below the scalar masses ($1$ TeV in the figure) or close to the $Z^\prime$ resonance region, as clearly seen in the figure.  Notice that $\Omega_{\rm DM}h^2$ in this case decreases with $\tan\beta$, allowing to satisfy the relic density constraint over a wide range of dark matter masses above the resonance. 

\begin{figure}[tb!]
\begin{center}
\includegraphics[scale=0.45]{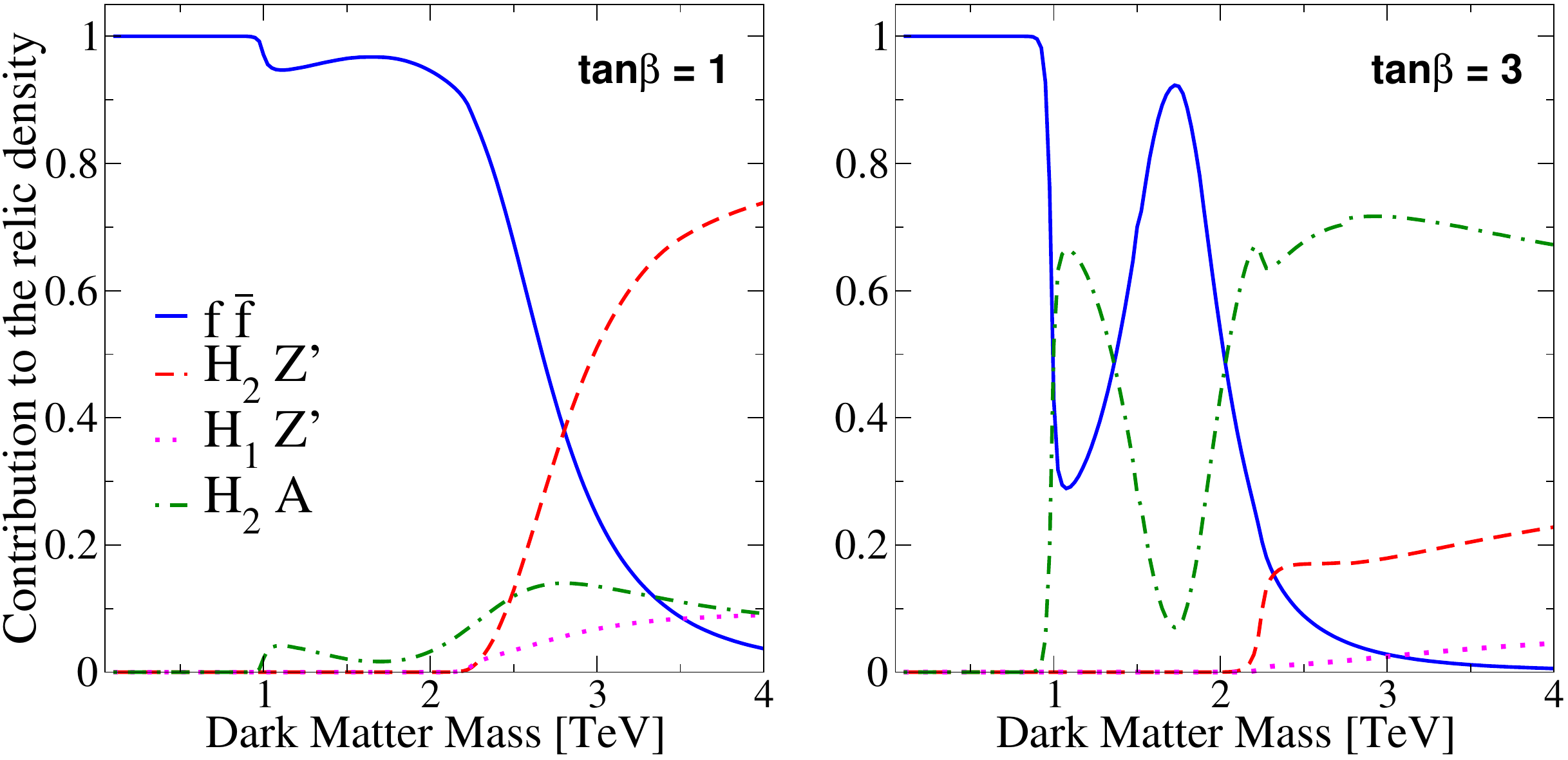}
\caption{\small The contributions from different final states to the relic density for  $\tan\beta=1$ (left panel) and $\tan\beta=3$ (right panel). The different lines in each panel correspond to  the most relevant final states: SM fermions (solid blue line), $H_2 Z^\prime$ (dashed red line),  $H_1Z^\prime$ (dotted magenta line) and $H_2 A_2$ (dash-dotted green line). In this figure, $\gbl=0.5$, $\mzp=3.5$ TeV, $\mhone=\mhtwo=M_A=1$ TeV and all mixing angles -- $\theta_{L,R}$, $\theta$ -- were set to zero. \label{fig:branchings}} 
\end{center}
\end{figure}

Figure \ref{fig:branchings} demonstrates how the contribution from different final states to the relic density changes with  the  dark matter mass and with $\tan\beta$.  It shows the ratio between the dark matter annihilation cross section into a given final state and the total annihilation cross section at freeze-out for the most relevant final states: SM fermions (solid blue line), $H_2 Z^\prime$ (dashed red line),  $H_1Z^\prime$ (dotted magenta line) and $H_2 A_2$ (dash-dotted green line). The left (right) panel corresponds to $\tan\beta=1$ ($\tan\beta=3$) while the rest of parameters are identical to those used in figure \ref{fig:rdtb}. Notice that, for $\tan\beta=1$, the annihilation into fermions is dominant up to a dark matter mass of about $3$ TeV. From then on, it is the $H_2Z^\prime$ final state that dominates. For $\tan\beta=3$ (right panel) the final state $H_2A$ becomes very important, dominating the annihilation rate for dark matter masses above $2$ TeV. In this case, the SM fermions constitute the primary annihilation channel for $\mdm<M_{H_{1,2},A}$   and also around the $Z^\prime$ resonance.

In the following, we will focus on the case where the $B-L$ gauge interactions, rather than the scalar ones, determine the relic density. This choice is motivated by several factors. On the one hand, such scenario is more predictive, because the  dark matter relic density depends only on four parameters: $\mdm$, $\mzp$, $\gbl$, and $\theta_{L}$. On the other hand, these gauge interactions also determine the expected signals in dark matter detection experiments  and  at the LHC, implying interesting correlations between different observables and providing a way to test this model in the near future.

\subsection{The viable parameter space}
\begin{table}[tb!]
\begin{center}
\begin{tabular}{c|c}
	\hline
	Parameter	& Range	\\
    \hline \hline
    $\mzp$ & $(200~\mathrm{GeV},50~\mathrm{TeV} )$\\
    $\mdm$ & $< \mzp$\\
    $M_{\psi_2}$ & $(1.2, 3.0)\mdm$\\
    $\gbl$ & $(0.001,1)$\\
    $\theta,\theta_{L,R}$ & $(0,2\pi)$\\
    $\tan\beta$ & $(0.03, 30)$\\
    $M_{H_1,H_2,A}$ & $(200~\mathrm{GeV},10~\mathrm{TeV})$\\
	\hline
\end{tabular}
\caption{\small Parameters of the model and their allowed range of variation in our scan.}
\end{center}
\label{tab:scan}
\end{table}

To determine the viable regions of this model, we have randomly scanned its parameter space within the ranges shown in Table \ref{tab:scan}. Then, we have selected those points  consistent with perturbativity ($|\lambda|<1$ for all dimensionless couplings $\lambda$), with the LEP II bound --equation \ref{eq:LEPII}-- and with the observed dark matter density, $\Omega_{\rm DM}h^2 \approx 0.12$ \cite{Ade:2013zuv}. As already stated, this relic density was assumed to be the result of a thermal freeze-out in the early Universe and to be dominated by the $B-L$ gauge interactions. The resulting  sample of points  represents what we call the viable parameter space of this model. Later, we will examine whether these viable points are also consistent with dark matter detection bounds and with current LHC searches. Let us first analyze this viable parameter space.

\begin{figure}[tb]
\begin{center}
\includegraphics[scale=0.4]{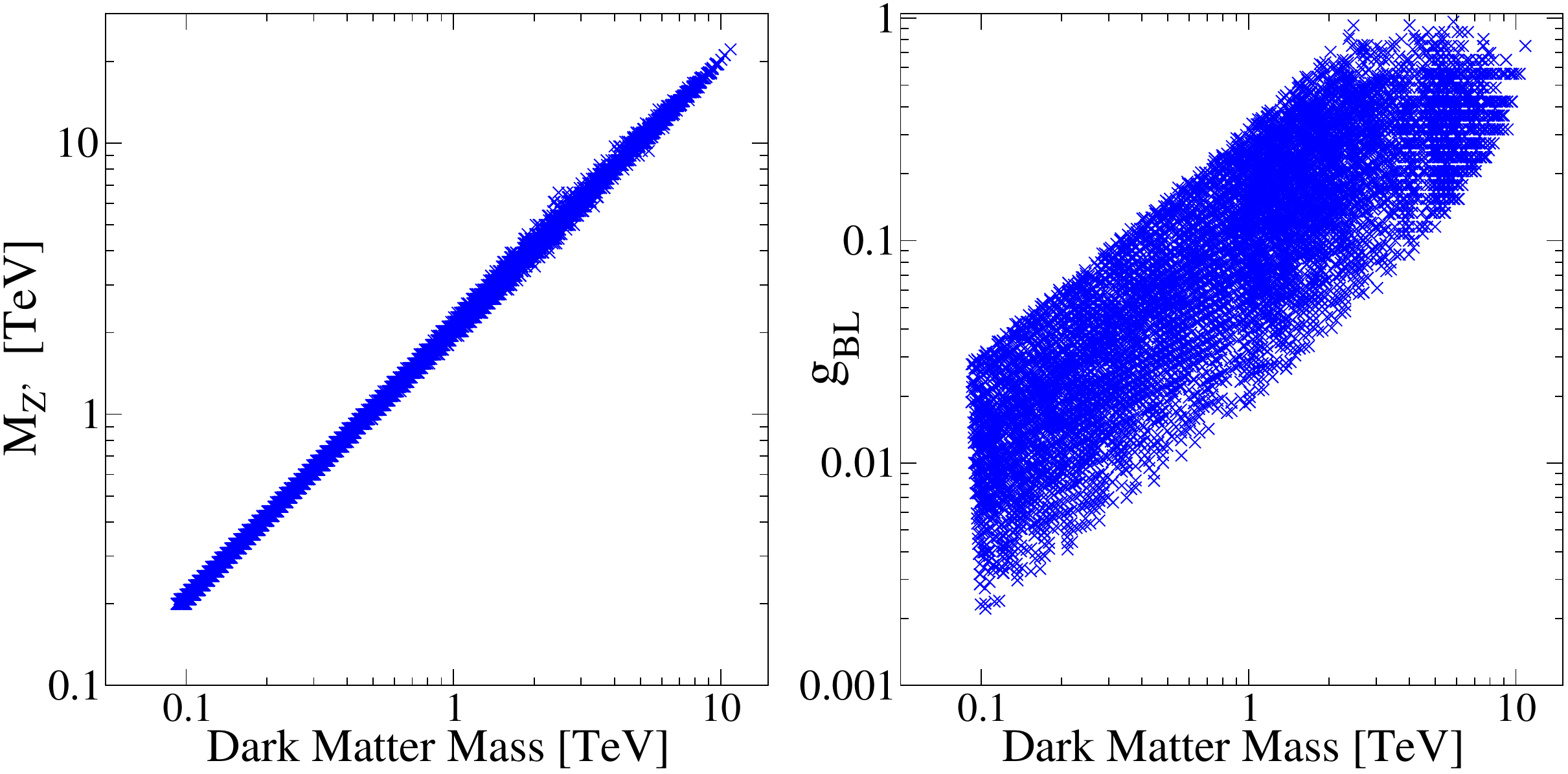}
\caption{\small The regions that are consistent with the dark matter constraint in the planes ($\mdm$, $\mzp$) and ($\mdm$, $\gbl$).   Dark matter annihilations are  assumed to be determined by the gauge interactions. \label{fig:scangauge}} 
\end{center}
\end{figure} 

Figure \ref{fig:scangauge} projects the viable parameter space onto two different planes: ($\mdm$, $\mzp$) in the left panel, and  ($\mdm$, $\gbl$) in the right one. From the left panel we see that the correct relic density can be  achieved over a wide range of dark matter masses, but always relatively close to the $Z^\prime$ resonance. Moreover, we obtain an upper limit on the dark matter mass of about $10$ TeV, and a corresponding upper limit on $\mzp$ of order $20$ TeV. From the right panel, we see that $\gbl$ can vary between $10^{-3}$, for light dark matter particles, and $1$ (the highest value allowed in our scan) for dark matter masses close to their upper limit. 

\begin{figure}[tb]
\begin{center}
\includegraphics[scale=0.4]{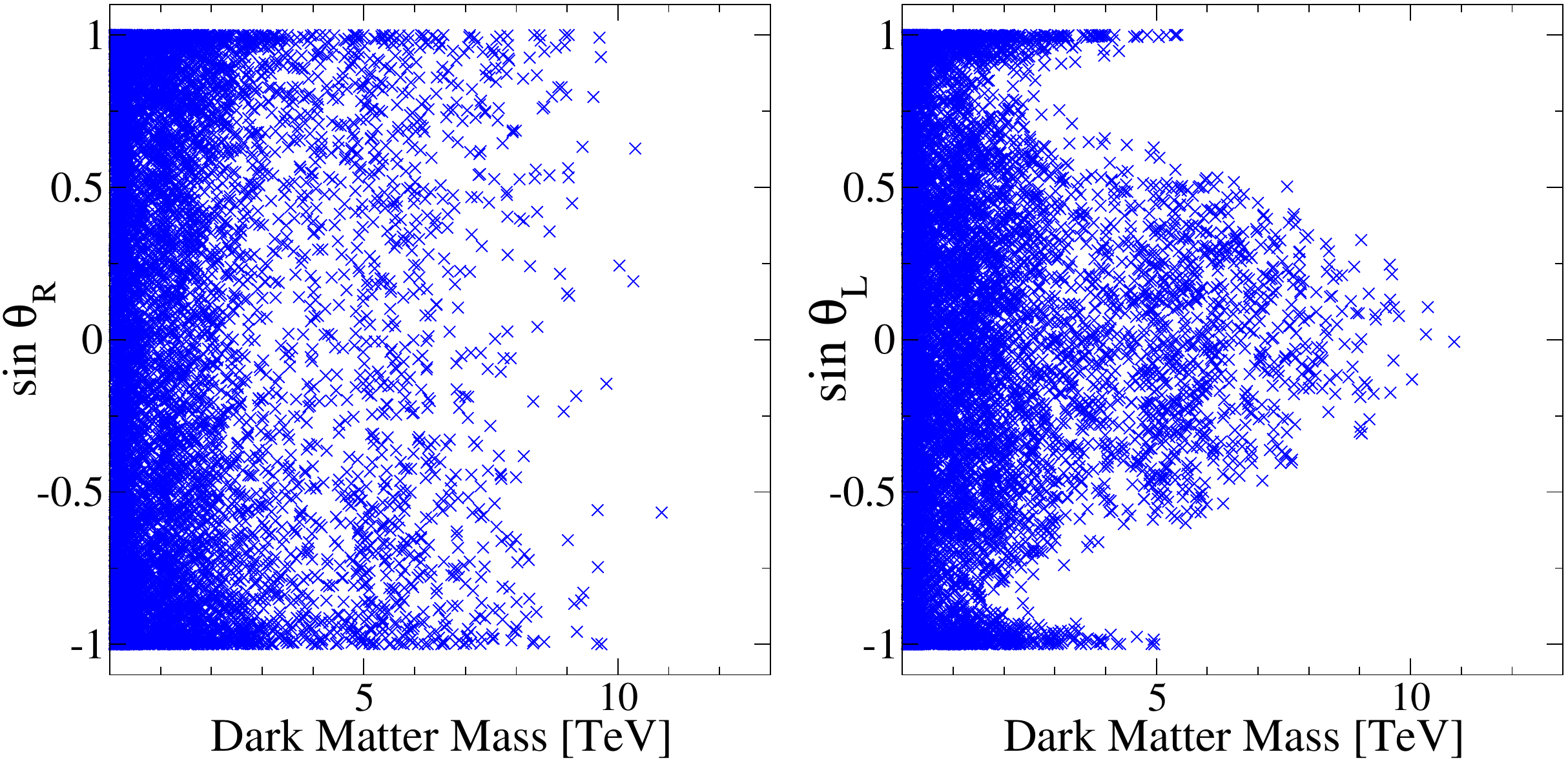}
\caption{\small The regions that are consistent with the dark matter constraint in the planes ($\mdm$, $\sin\theta_R$) and ($\mdm$, $\sin\theta_L$).   \label{fig:scangaugeangles}} 
\end{center}
\end{figure} 

The other parameters that could affect the relic density are the fermion mixing angles, $\theta_{L,R}$. We already saw, however, that only $\theta_L$ enters into the dark matter annihilation cross section. The left panel of figure \ref{fig:scangaugeangles} shows that indeed the viable points have no preference for any particular value of $\theta_R$. A different pattern emerges for $\theta_L$, as illustrated in the right panel. In this case, we see that at high dark matter masses $\theta_L$ has to be small. This result is in agreement with the fact that for small values of $\theta_L$, the left-handed component of the dark matter particle coincides with the field $\xi_L$, which has the largest $B-L$
 charge among the new fermions. Alternatively, one can see that $g_{\psi V}$ is maximized for $\theta_L=0$. Notice that for $\sin\theta_L\approx\pm 1$, corresponding to $\psi_{1L}\approx \eta_L$, the upper limit on the dark matter mass is smaller, of order $6$ TeV. 
 
We have now characterized the viable regions of this model, those consistent with the LEP limit and with the observed dark matter density. It remains to be seen whether these regions are also compatible with  current limits from  direct and indirect dark matter detection experiments, and whether they can be probed in future experiments.

\subsection{Indirect detection}
\begin{figure}[tb]
\begin{center}
\includegraphics[scale=0.4]{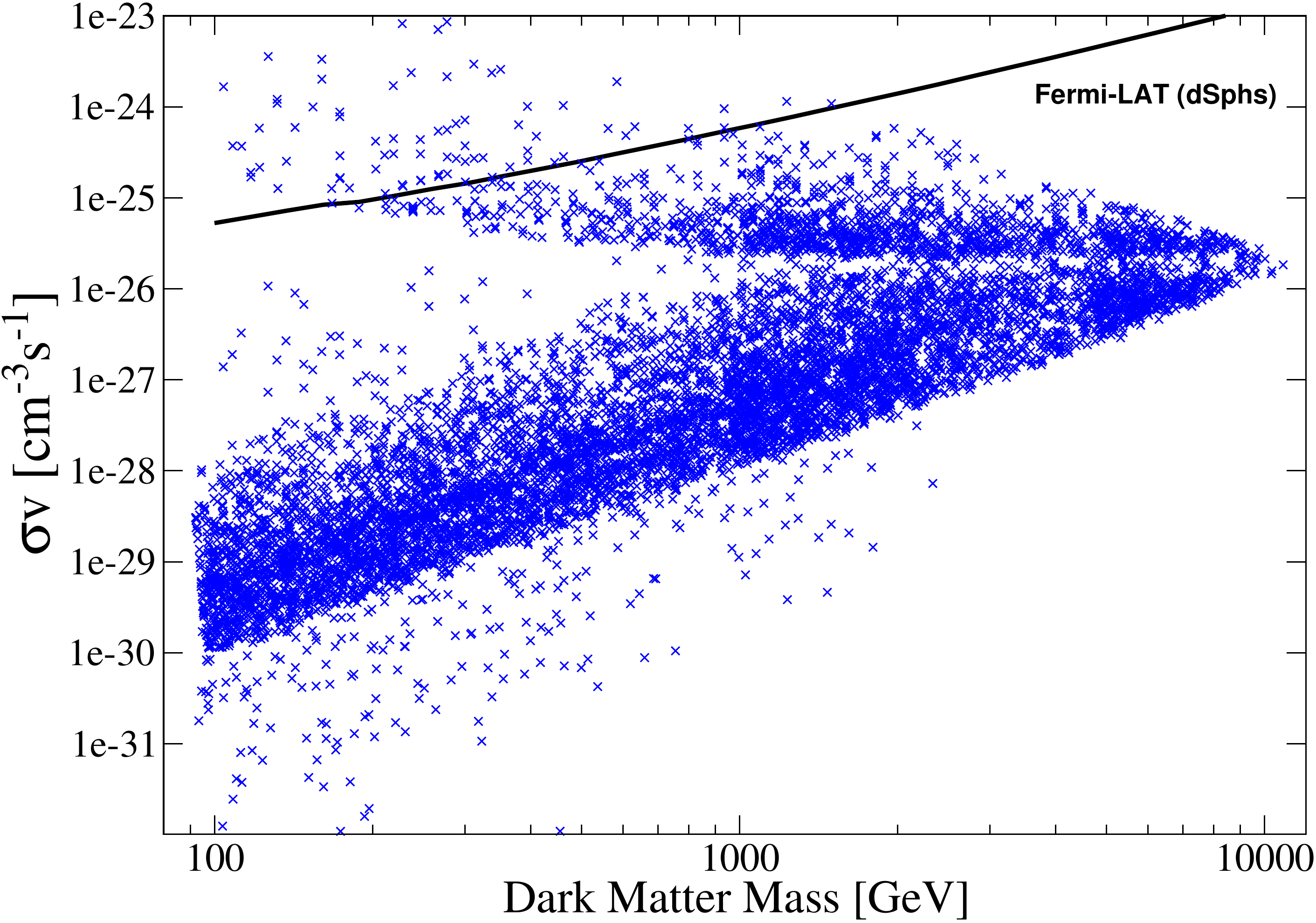}
\caption{\small The dark matter annihilation cross section versus the dark matter mass for our sample of viable models. The solid black line shows the   limit we have derived for this particular model from Fermi-LAT data. This limit is based on the  stacked analysis of 15 dwarf spheroidal galaxies presented in \cite{Ackermann:2015zua}.  \label{fig:scangaugesv}} 
\end{center}
\end{figure} 

The dark matter indirect detection signals -- $\gamma$, $\nu$, $e^+$ and $\bar p$ fluxes -- are determined, on the particle physics side, by the mass of the dark matter particle, its annihilation cross section $\sigma v$, and the annihilation final states.  The differential photon flux  from a given angular direction $\Delta\Omega$, for example, can be expressed as
\begin{equation}
\frac{d\Phi_\gamma(\Delta \Omega)}{dE}(E_\gamma)=\frac{1}{4\pi}\frac{\sigma v}{2 M_{DM}^2}\sum_i Br_i\frac{dN^i_\gamma}{dE_{\gamma}}\cdot J_{\rm ann}\;,
\end{equation}
where the index $i$ runs over the different final states from dark matter annihilation,  $Br_i$ is the branching ratio into the $i$ final state, and $\frac{dN^i_\gamma}{dE_{\gamma}}$ is the differential $\gamma$-ray yield per annihilation into the $i$ final state;  $J_{\rm ann}$ is instead the annihilation $J$-factor, which characterizes the astrophysical environment (e.g. the galactic center or a dwarf galaxy) where the signal is produced.

Figure \ref{fig:scangaugesv} displays our set of viable models in the plane $\mdm$ versus $\sigma v$. Since the relic density constraint in this model is always satisfied close to the $Z^\prime$ resonance ($\mdm\sim \mzp/2$), the usual argument for a \emph{thermal}  annihilation cross section of order $\sim 3\times 10^{-26}\,\mathrm{cm}^{-3}\, \mathrm{s}^{-1}$ does not apply. Instead, we see that most points feature smaller cross sections, with some reaching values even below 
$10^{-30}\,\mathrm{cm}^{-3}\, \mathrm{s}^{-1}$. Such small cross sections are very challenging for indirect detection experiments. Notice, however,  that the minimum value of $\sigma v$ increases with the dark matter mass, lying above $10^{-27}\,\mathrm{cm}^{-3}\, \mathrm{s}^{-1}$ for $\mdm\gtrsim 3$ TeV. Other points feature instead larger cross sections, with values as high as  $10^{-23}\,\mathrm{cm}^{-3}\, \mathrm{s}^{-1}$ for low dark matter masses. The maximum value of $\sigma v$ is also observed to decrease with the dark matter mass, lying  below $10^{-25}\,\mathrm{cm}^{-3}\, \mathrm{s}^{-1}$ for $\mdm\gtrsim 4$ TeV.

Currently, the Fermi-LAT limits from dwarf spheroidal galaxies \cite{Ackermann:2015zua} provide the strongest constraints on the annihilation cross section  over a wide range of dark matter masses. These constraints are presented separately, in the plane dark matter mass versus $\sigma v$, for different annihilation channels ($\ell^+\ell^-$, $q\bar q$, $W^+W^-$, $b\bar b$), assuming a $100\%$ branching ratio in each case. Thus, they can be directly applied to models where a single channel tends to dominate the annihilation cross section. That is not the case in our model, however, because being a process mediated by the $Z^\prime$, the dark matter annihilates into all SM fermions with comparable branching ratios. In fact, each fermion contributes to the annihilation rate with a weigth proportional to $N_c Q_{B-L}^2$, where $N_c$ is its color factor and $Q_{B-L}$  its $B-L$ charge.  We have used these branchings and the likelihoods provided by the Fermi-LAT collaboration \cite{Fermidata} to derive an upper limit on $\sigma v$ (at $95\%$ CL) for our model (valid also for other models based on the $B-L$ gauge symmetry). 
This limit is displayed in figure \ref{fig:scangaugesv} as a solid black line. 
Notice  that only few points  are currently  excluded by the Fermi-LAT data.  All of them feature  dark matter masses below $1.5$ TeV and annihilation cross sections higher than the  thermal one. In the near future, the Cherenkov Telescope Array (CTA) \cite{Acharya:2013sxa} is expected to significantly improve the Fermi-LAT limits, particularly at high dark matter masses.  But given the suppressed value of $\sigma v$ that is typical of this model, the impact of the CTA on the viable parameter space is expected to be very limited.

\subsection{Direct detection}
\begin{figure}[tb]
\begin{center}
\includegraphics[scale=0.4]{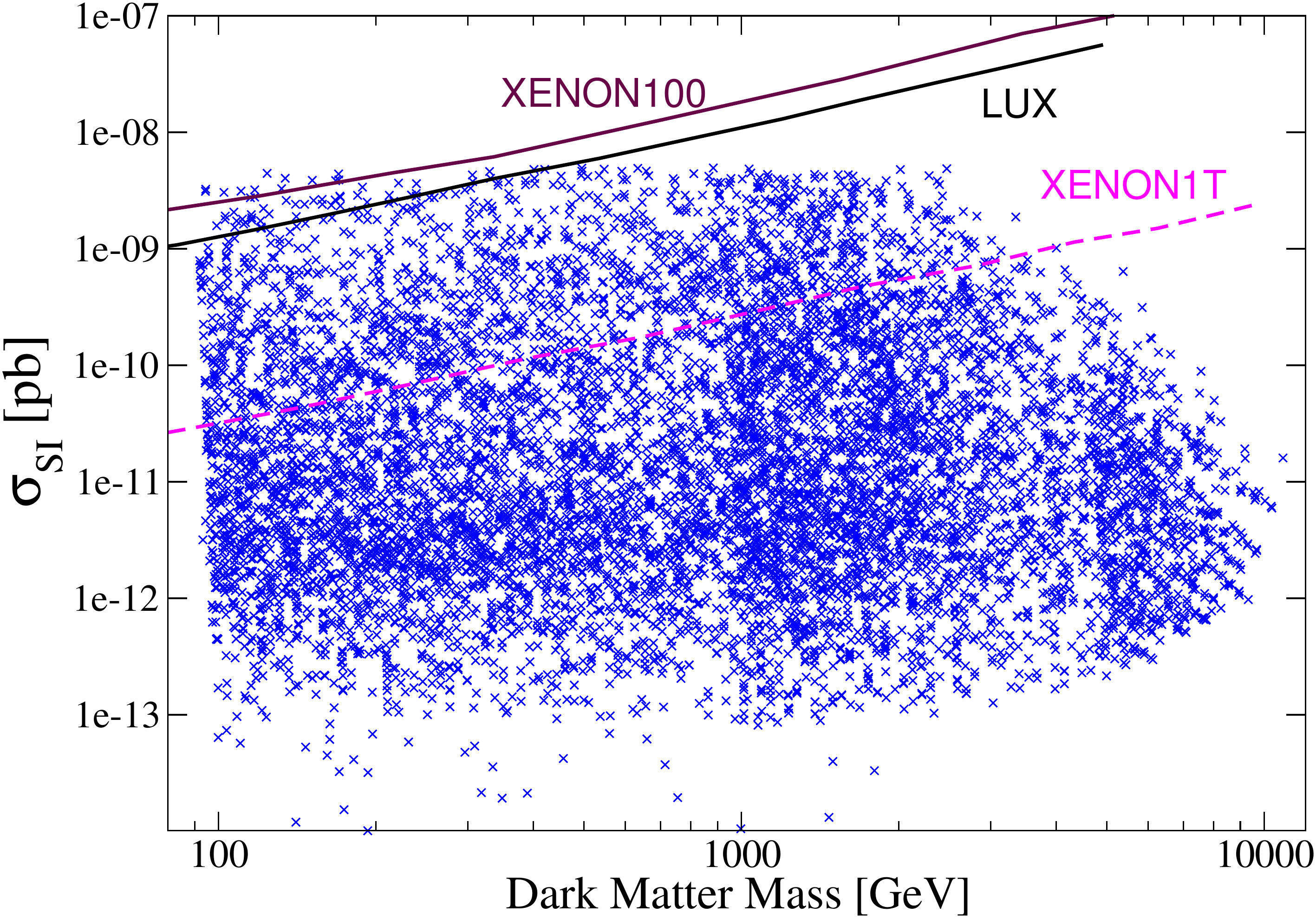}
\caption{\small The predicted spin-independent cross section versus the dark matter mass for models consistent with the dark matter constraint.  For illustration, the current limits from XENON100 and LUX are also shown (solid lines) as well as the expected sensitivity of XENON1T (dashed line). The relic density is assumed to be determined by the gauge interactions. \label{fig:scangaugesip}} 
\end{center}
\end{figure} 
The dark matter particle in this model can scatter coherently off nuclei  via a tree-level $Z^\prime$-exchange diagram. Since the coupling between the quarks and the $Z^\prime$ is vector-like, the spin-dependent cross section vanishes and only the spin-independent (SI) one contributes to the scattering. The dark matter-nucleon SI cross section is given by
\begin{eqnarray}
\sigma_{\rm SI} \simeq \frac{m_n^2 }{4 \pi}\frac{\gbl^2\, g_{\psi V}^2} {\mzp^4}, 
\end{eqnarray}
where we use the fact that the nucleon mass, $m_n$, is  much smaller than the dark matter mass. The direct detection cross section of the dark matter is thus determined by just three parameters: $\gbl$, $\mzp$, and $\theta_L$.

Figure \ref{fig:scangaugesip} shows the spin-independent cross section versus the dark matter mass for the viable parameter space. We see that the predicted cross section varies approximately between $10^{-13}$ and $10^{-8}$ pb. Current limits from XENON100~\cite{Aprile:2012nq} and LUX~\cite{Akerib:2013tjd} are  displayed as solid lines. They can  exclude only a handful of points with dark matter masses below $400$ GeV. The expected sensitivity of the  XENON1T experiment ~\cite{Aprile:2015uzo}, which is already taking data, is also shown as a dashed line. From the figure we see that XENON1T will probe dark matter masses as high as $4$ TeV and may exclude a significant fraction of the viable parameter space. 

\section{LHC bounds}
\label{sec:lhc}

\begin{figure}[h!]
\begin{center}
\includegraphics[scale=0.4]{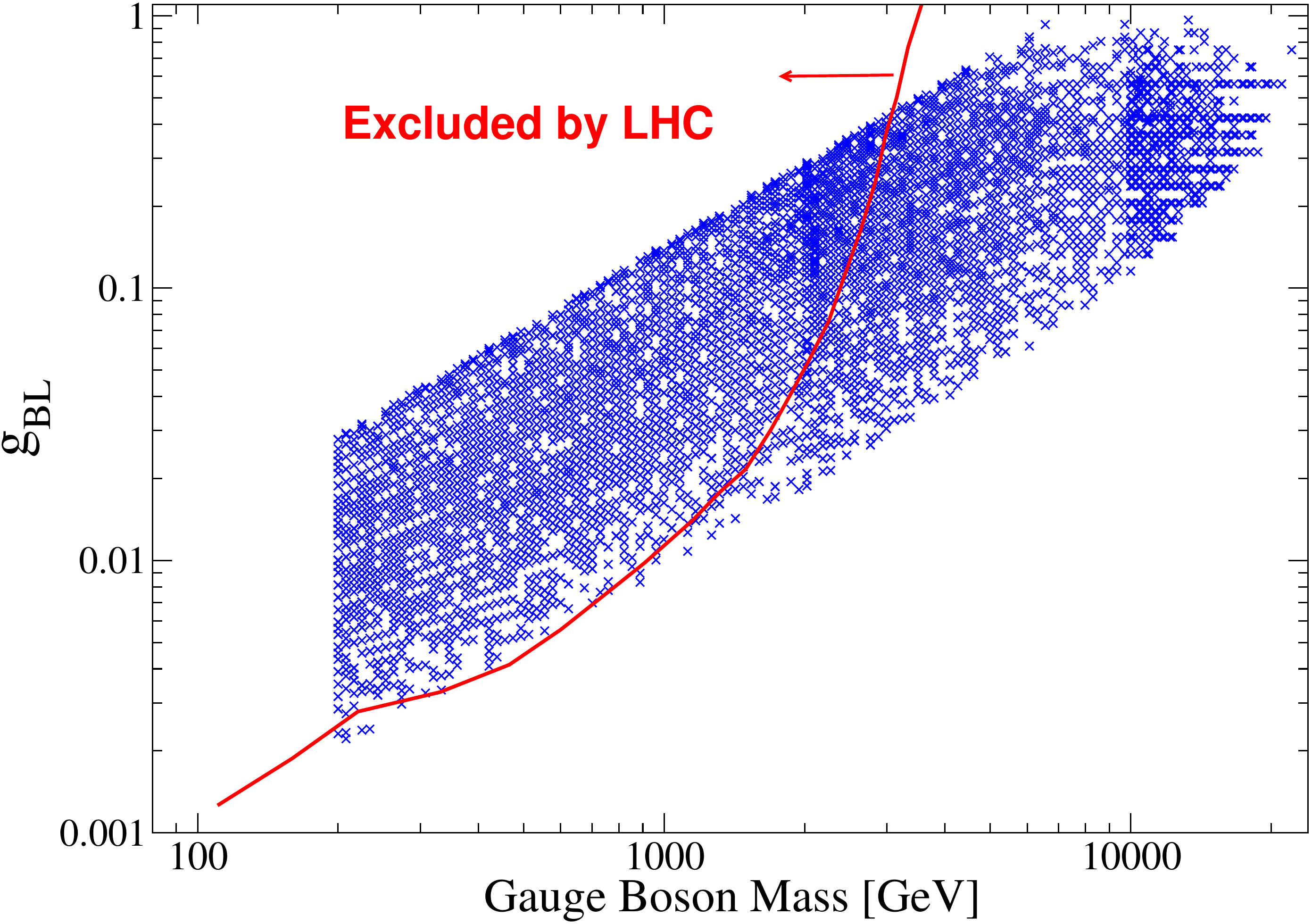}
\caption{\small The LHC exclusion regions in the plane ($\mzp$, $\gbl$). The points to the left of the red line are already ruled out by dilepton searches at the LHC. Notice that the region $\mzp\lesssim 1$ TeV (or $\mdm\lesssim 500$ GeV) is essentially excluded.   \label{fig:scangaugelhc}} 
\end{center}
\end{figure}

The $Z^\prime$ boson of the gauged $B-L$ model can be produced at  hadron colliders \cite{Carena:2004xs} such as the LHC. Current dilepton limits ($q\bar q\to Z^\prime\to \ell\bar{\ell}$) from the LHC \cite{Aad:2014cka}, in fact, provide stringent constraints on the $B-L$ gauge boson, as recently shown  in \cite{Guo:2015lxa}.  Figure \ref{fig:scangaugelhc} compares, in the plane ($\mdm$, $\gbl$),  our viable parameter space with the LHC limits. Points lying to the left of the red line are not consistent with the LHC observations.  We see that the region $\mzp\lesssim 1$ TeV (corresponding to a dark matter mass $\mdm\lesssim 500$ GeV, see figure \ref{fig:scangauge}) is completely excluded by LHC data. For $Z^\prime$ masses between $1$ and $3$ TeV, some points are ruled out, depending on the value of $\gbl$.  The region $\mzp\gtrsim 3$ TeV, on the other hand, is not constrained by current data. In the near future, these limits will become more stringent as more data from the 13 TeV run becomes available.

%

\section{Neutrino masses}
\label{sec:numass}
Since the right-handed neutrinos are absent in our model, neutrino masses cannot be generated via the type-I seesaw mechanism. A simple alternative seems to be the type-II seesaw \cite{Schechter:1980gr,Magg:1980ut,Mohapatra:1980yp,Lazarides:1980nt}, in which  a scalar triplet $\Delta$ is added to the SM and gets, after electroweak symmetry breaking,  a small vev induced by the trilinear term $\mu_\Delta \Delta H H$.  The interaction term $f_\Delta \ell_L^T \Delta \ell_L$ ($\ell_L$ being the left-handed SM fermion doublet) then generates light neutrino masses of the form $m_\nu=f_\Delta \langle \Delta \rangle$. The smallness of neutrino masses is explained in this case by the fact that $\langle\Delta\rangle\propto 1/M_{\Delta}^2$, which is suppressed for a heavy $\Delta$. This type-II seesaw is then the motivation to include a scalar triplet in the particle content of our model -- see section \ref{sec:model}.

Notice, though, that the extra $B-L$ symmetry of our model prevents the realization of the standard type-II seesaw as mentioned above. In fact, the coupling $f_\Delta \ell_L^T \Delta \ell_L$ fixes the  $B-L$  number of $\Delta$ to be $-2$, which implies that the term $\mu_\Delta \Delta H H$ is  forbidden -- since $H$ has $B-L$ equal to zero. Remarkably, the field $\phi_2$, which is already part of this model,  provides a way out of this problem, as we now explain.

\begin{figure}[tb!]
	\centering
	\includegraphics[width=0.75\textwidth]{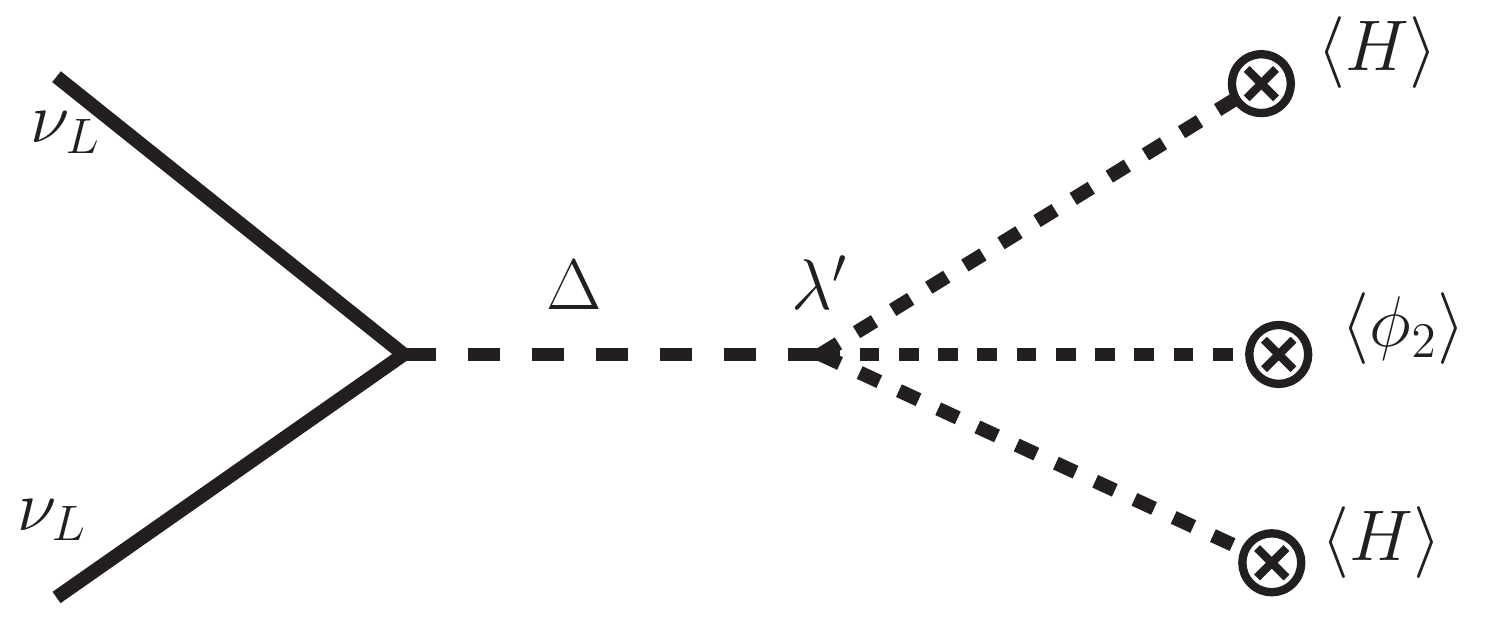}
	\caption{\small Diagrammatic representation for the neutrino mass generation in this model. }\label{fig:numass}
\end{figure}

The  Lagrangian terms involving the triplet field are given by  
\begin{align}
\mathcal{L}_\Delta =& 
\mbox{Tr}\left[\left(\mathcal{D}_\mu \Delta \right)^\dagger \left(\mathcal{D}^\mu \Delta \right) \right] 
-\bigg[ M^2_\Delta \mbox{Tr}\left[\Delta^\dagger \Delta \right]
+ \lambda_\Delta \mbox{Tr}\left[\Delta^\dagger \Delta \right]^2\\
&+\lambda_{\Delta H} \mbox{Tr}\left[\Delta^\dagger \Delta \right] (H^\dagger H)
+\lambda_{\Delta \phi_1} \mbox{Tr}\left[\Delta^\dagger \Delta \right] (\phi_1^\dagger \phi_1) \nonumber \\
&+\lambda_{\Delta \phi_2} \mbox{Tr}\left[\Delta^\dagger \Delta \right] (\phi_2^\dagger \phi_2)
+\lambda^\prime H^T i \tau_2 \Delta H \phi_2 \bigg] 
+ f_{ij} \ell^T_{iL} \mathcal{C} \sigma_2 \Delta \ell_{jL} + h.c., \nonumber
\end{align}
with $\mathcal{D}_\mu= \partial_{\mu}  + i\,g_L T^a W^a_{\mu} + 2 i\,g_\text{BL} \,Z_\mu^\prime$.  We see that the term $H \Delta H \phi_2$ could play the role of the  trilinear term $\mu_\Delta \Delta H H$ in the standard type-II seesaw. The dimensionful term $\mu_\Delta$ is here replaced by the scalar field $\phi_2$ (see figure \ref{fig:numass}), which is also responsible for the breaking of $B-L$ and the  masses of the new fermions.  Thus, the neutrino mass in this model is indirectly linked  to the $B-L$ breaking scale and the dark matter mass. 
The induced vev of the triplet is given in this model by 
\begin{align}
v_\Delta = \langle \Delta \rangle \simeq \frac{\lambda^\prime v^2 v_2}{M^2_\Delta} \, , 
\end{align}
and  the light neutrino mass matrix becomes 
\begin{align}
m_\nu= f \langle \Delta \rangle \simeq f \frac{\lambda^\prime v^2 v_2}{M^2_\Delta} \,.
\end{align}
For typical input model parameters $v\simeq 100~$GeV, $v_2 \simeq~$TeV and $\lambda^\prime \leq {\cal O}(1)$,  the ${\cal O}(0.1)~$eV scale of light neutrinos with Yukawa couplings $f\simeq 0.1$ is generated by  scalar triplet masses around $10^{8}~$GeV.

\section{Conclusions}
\label{sec:con}
We presented a new $U(1)_{B-L}$ gauge extension of the Standard Model in which the gauge anomalies are canceled by four chiral fermions with fractional $B-L$ charges rather than by  right-handed neutrinos. The resulting scenario is  simple and predictive and can simultaneously explain  dark matter and neutrino masses. A remarkable  feature of this framework is that the lightest among the new  fermions originally introduced for  anomaly cancellation is automatically  stable -- the $B-L$ charges forbid any tree level interaction with the Standard Model particles -- and constitutes a viable dark matter candidate.  We determined the regions of the parameter space that are consistent with the observed dark matter density and showed that they can be partially tested in current and future dark matter experiments. LHC searches were also found to constrain the parameter space in an important way. Finally, we showed that neutrino masses in this model can be explained by a type-II seesaw mechanism involving also one of the fields responsible for the dark matter mass. This new $B-L$ model is thus an attractive and testable scenario for physics beyond the Standard Model.

\section*{Acknowledgements} 
Work supported by the DFG in the Heisenberg programme with grant 
RO 2516/6-1 (WR), and by the Max Planck Society in the project 
MANITOP (WR, CY). The work of SP is partially supported by the Department 
of Science and Technology, Govt.\ of India under the financial 
grant SB/S2/HEP-011/2013. 

\bibliographystyle{hunsrt}
\bibliography{darkmatter}

\begin{thebibliography}{10}

\bibitem{Jenkins:1987ue}
Elizabeth~Ellen Jenkins.
\newblock {Searching for a B-L Gauge Boson in $p \bar{p}$ Collisions}.
\newblock {\em Phys. Lett.}, B192:219--222, 1987.

\bibitem{Buchmuller:1991ce}
W.~Buchmuller, C.~Greub, and P.~Minkowski.
\newblock {Neutrino masses, neutral vector bosons and the scale of B-L
  breaking}.
\newblock {\em Phys. Lett.}, B267:395--399, 1991.

\bibitem{Khalil:2006yi}
Shaaban Khalil.
\newblock {Low scale $B$ - L extension of the Standard Model at the LHC}.
\newblock {\em J. Phys.}, G35:055001, 2008, hep-ph/0611205.

\bibitem{Basso:2008iv}
Lorenzo Basso, Alexander Belyaev, Stefano Moretti, and Claire~H.
  Shepherd-Themistocleous.
\newblock {Phenomenology of the minimal B-L extension of the Standard model: Z'
  and neutrinos}.
\newblock {\em Phys. Rev.}, D80:055030, 2009, 0812.4313.

\bibitem{Iso:2009ss}
Satoshi Iso, Nobuchika Okada, and Yuta Orikasa.
\newblock {Classically conformal $B^-$ L extended Standard Model}.
\newblock {\em Phys. Lett.}, B676:81--87, 2009, 0902.4050.

\bibitem{Kanemura:2014rpa}
Shinya Kanemura, Toshinori Matsui, and Hiroaki Sugiyama.
\newblock {Neutrino mass and dark matter from gauged $U(1)_{B-L}$ breaking}.
\newblock {\em Phys. Rev.}, D90:013001, 2014, 1405.1935.

\bibitem{Petcov:2013poa}
S.~T. Petcov.
\newblock {The Nature of Massive Neutrinos}.
\newblock {\em Adv. High Energy Phys.}, 2013:852987, 2013, 1303.5819.

\bibitem{Gonzalez-Garcia:2015qrr}
M.~C. Gonzalez-Garcia, Michele Maltoni, and Thomas Schwetz.
\newblock {Global Analyses of Neutrino Oscillation Experiments}.
\newblock {\em Nucl. Phys.}, B908:199--217, 2016, 1512.06856.

\bibitem{Bergstrom:2000pn}
Lars Bergström.
\newblock {Nonbaryonic dark matter: Observational evidence and detection
  methods}.
\newblock {\em Rept. Prog. Phys.}, 63:793, 2000, hep-ph/0002126.

\bibitem{Bertone:2004pz}
Gianfranco Bertone, Dan Hooper, and Joseph Silk.
\newblock {Particle dark matter: Evidence, candidates and constraints}.
\newblock {\em Phys. Rept.}, 405:279--390, 2005, hep-ph/0404175.

\bibitem{Drees:2012ji}
Manuel Drees and Gilles Gerbier.
\newblock {Mini-Review of Dark Matter: 2012}.
\newblock 2012, 1204.2373.

\bibitem{Minkowski:1977sc}
Peter Minkowski.
\newblock {$\mu \to e\gamma$ at a Rate of One Out of $10^{9}$ Muon Decays?}
\newblock {\em Phys. Lett.}, B67:421--428, 1977.

\bibitem{Yanagida:1979as}
Tsutomu Yanagida.
\newblock {HORIZONTAL SYMMETRY AND MASSES OF NEUTRINOS}.
\newblock {\em Conf. Proc.}, C7902131:95--99, 1979.

\bibitem{GellMann:1980vs}
Murray Gell-Mann, Pierre Ramond, and Richard Slansky.
\newblock {Complex Spinors and Unified Theories}.
\newblock {\em Conf. Proc.}, C790927:315--321, 1979, 1306.4669.

\bibitem{Mohapatra:1979ia}
Rabindra~N. Mohapatra and Goran Senjanovic.
\newblock {Neutrino Mass and Spontaneous Parity Violation}.
\newblock {\em Phys. Rev. Lett.}, 44:912, 1980.

\bibitem{Rodejohann:2015lca}
Werner Rodejohann and Carlos~E. Yaguna.
\newblock {Scalar dark matter in the B−L model}.
\newblock {\em JCAP}, 1512(12):032, 2015, 1509.04036.

\bibitem{Lindner:2013awa}
Manfred Lindner, Daniel Schmidt, and Atsushi Watanabe.
\newblock {Dark matter and symmetry for the right-handed neutrinos}.
\newblock {\em Phys. Rev.}, D89(1):013007, 2014, 1310.6582.

\bibitem{Okada:2010wd}
Nobuchika Okada and Osamu Seto.
\newblock {Higgs portal dark matter in the minimal gauged $U(1)_{B-L}$ model}.
\newblock {\em Phys. Rev.}, D82:023507, 2010, 1002.2525.

\bibitem{Okada:2012sg}
Nobuchika Okada and Yuta Orikasa.
\newblock {Dark matter in the classically conformal B-L model}.
\newblock {\em Phys. Rev.}, D85:115006, 2012, 1202.1405.

\bibitem{Basak:2013cga}
Tanushree Basak and Tanmoy Mondal.
\newblock {Constraining Minimal $U(1)_{B-L}$ model from Dark Matter
  Observations}.
\newblock {\em Phys. Rev.}, D89:063527, 2014, 1308.0023.

\bibitem{Duerr:2015wfa}
Michael Duerr, Pavel Fileviez~Perez, and Juri Smirnov.
\newblock {Simplified Dirac Dark Matter Models and Gamma-Ray Lines}.
\newblock {\em Phys. Rev.}, D92(8):083521, 2015, 1506.05107.

\bibitem{Guo:2015lxa}
Jun Guo, Zhaofeng Kang, P.~Ko, and Yuta Orikasa.
\newblock {Accidental dark matter: Case in the scale invariant local B-L
  model}.
\newblock {\em Phys. Rev.}, D91(11):115017, 2015, 1502.00508.

\bibitem{Dasgupta:2016odo}
Arnab Dasgupta, Chandan Hati, Sudhanwa Patra, and Utpal Sarkar.
\newblock {A minimal model of TeV scale WIMPy leptogenesis}.
\newblock 2016, 1605.01292.

\bibitem{Montero:2007cd}
J.~C. Montero and V.~Pleitez.
\newblock {Gauging U(1) symmetries and the number of right-handed neutrinos}.
\newblock {\em Phys. Lett.}, B675:64--68, 2009, 0706.0473.

\bibitem{Sanchez-Vega:2014rka}
B.~L. Sánchez-Vega, J.~C. Montero, and E.~R. Schmitz.
\newblock {Complex Scalar DM in a B-L Model}.
\newblock {\em Phys. Rev.}, D90(5):055022, 2014, 1404.5973.

\bibitem{Ma:2014qra}
Ernest Ma and Rahul Srivastava.
\newblock {Dirac or inverse seesaw neutrino masses with $B-L$ gauge symmetry
  and $S_3$ flavor symmetry}.
\newblock {\em Phys. Lett.}, B741:217--222, 2015, 1411.5042.

\bibitem{Sanchez-Vega:2015qva}
B.~L. Sánchez-Vega and E.~R. Schmitz.
\newblock {Fermionic dark matter and neutrino masses in a B-L model}.
\newblock {\em Phys. Rev.}, D92:053007, 2015, 1505.03595.

\bibitem{Ma:2015mjd}
Ernest Ma, Nicholas Pollard, Rahul Srivastava, and Mohammadreza Zakeri.
\newblock {Gauge $B-L$ Model with Residual $Z_3$ Symmetry}.
\newblock {\em Phys. Lett.}, B750:135--138, 2015, 1507.03943.

\bibitem{Schechter:1980gr}
J.~Schechter and J.~W.~F. Valle.
\newblock {Neutrino Masses in SU(2) x U(1) Theories}.
\newblock {\em Phys. Rev.}, D22:2227, 1980.

\bibitem{Magg:1980ut}
M.~Magg and C.~Wetterich.
\newblock {Neutrino Mass Problem and Gauge Hierarchy}.
\newblock {\em Phys. Lett.}, B94:61--64, 1980.

\bibitem{Mohapatra:1980yp}
Rabindra~N. Mohapatra and Goran Senjanovic.
\newblock {Neutrino Masses and Mixings in Gauge Models with Spontaneous Parity
  Violation}.
\newblock {\em Phys. Rev.}, D23:165, 1981.

\bibitem{Lazarides:1980nt}
George Lazarides, Q.~Shafi, and C.~Wetterich.
\newblock {Proton Lifetime and Fermion Masses in an SO(10) Model}.
\newblock {\em Nucl. Phys.}, B181:287--300, 1981.

\bibitem{Carena:2004xs}
Marcela Carena, Alejandro Daleo, Bogdan~A. Dobrescu, and Timothy M.~P. Tait.
\newblock {$Z^\prime$ gauge bosons at the Tevatron}.
\newblock {\em Phys. Rev.}, D70:093009, 2004, hep-ph/0408098.

\bibitem{Cacciapaglia:2006pk}
G.~Cacciapaglia, C.~Csaki, G.~Marandella, and A.~Strumia.
\newblock {The Minimal Set of Electroweak Precision Parameters}.
\newblock {\em Phys. Rev.}, D74:033011, 2006, hep-ph/0604111.

\bibitem{Khachatryan:2014jba}
Vardan Khachatryan et~al.
\newblock {Precise determination of the mass of the Higgs boson and tests of
  compatibility of its couplings with the standard model predictions using
  proton collisions at 7 and 8 $\,\text {TeV}$}.
\newblock {\em Eur. Phys. J.}, C75(5):212, 2015, 1412.8662.

\bibitem{Aad:2015zhl}
Georges Aad et~al.
\newblock {Combined Measurement of the Higgs Boson Mass in $pp$ Collisions at
  $\sqrt{s}=7$ and 8 TeV with the ATLAS and CMS Experiments}.
\newblock {\em Phys. Rev. Lett.}, 114:191803, 2015, 1503.07589.

\bibitem{Belanger:2013oya}
G.~Belanger, F.~Boudjema, A.~Pukhov, and A.~Semenov.
\newblock {micrOMEGAs 3: A program for calculating dark matter observables}.
\newblock {\em Comput. Phys. Commun.}, 185:960--985, 2014, 1305.0237.

\bibitem{Semenov:2010qt}
A.~Semenov.
\newblock {LanHEP - a package for automatic generation of Feynman rules from
  the Lagrangian. Updated version 3.1}.
\newblock 2010, 1005.1909.

\bibitem{Ade:2013zuv}
P.~A.~R. Ade et~al.
\newblock {Planck 2013 results. XVI. Cosmological parameters}.
\newblock {\em Astron. Astrophys.}, 571:A16, 2014, 1303.5076.

\bibitem{Ackermann:2015zua}
M.~Ackermann et~al.
\newblock {Searching for Dark Matter Annihilation from Milky Way Dwarf
  Spheroidal Galaxies with Six Years of Fermi Large Area Telescope Data}.
\newblock {\em Phys. Rev. Lett.}, 115(23):231301, 2015, 1503.02641.

\bibitem{Fermidata}
Figures and data files for a lat paper.
\newblock \url{http://www-glast.stanford.edu/pub_data/1048/}.

\bibitem{Acharya:2013sxa}
B.~S. Acharya et~al.
\newblock {Introducing the CTA concept}.
\newblock {\em Astropart. Phys.}, 43:3--18, 2013.

\bibitem{Aprile:2012nq}
E.~Aprile et~al.
\newblock {Dark Matter Results from 225 Live Days of XENON100 Data}.
\newblock {\em Phys.Rev.Lett.}, 109:181301, 2012, 1207.5988.

\bibitem{Akerib:2013tjd}
D.~S. Akerib et~al.
\newblock {First results from the LUX dark matter experiment at the Sanford
  Underground Research Facility}.
\newblock {\em Phys. Rev. Lett.}, 112:091303, 2014, 1310.8214.

\bibitem{Aprile:2015uzo}
E.~Aprile et~al.
\newblock {Physics reach of the XENON1T dark matter experiment}.
\newblock {\em JCAP}, 1604(04):027, 2016, 1512.07501.

\bibitem{Aad:2014cka}
Georges Aad et~al.
\newblock {Search for high-mass dilepton resonances in pp collisions at
  $\sqrt{s}=8$  TeV with the ATLAS detector}.
\newblock {\em Phys. Rev.}, D90(5):052005, 2014, 1405.4123.

\end{thebibliography}

\end{document}